\documentstyle{AGILEpsr}

\input epsf.sty
\input epsfig.sty
\input psfig.sty

\def \AAP #1 #2 {{\em Astron. Astrophys.\/} {\bf #1}, #2}
\def \AAL #1 #2 {{\em Astron. Astrophys. Lett.\/} {\bf #1}, L#2}
\def \AAR #1 #2 {{\em Astron. Astrophys. Rev.\/} {\bf #1}, #2}
\def \AAS #1 #2 {{\em Astron. Astrophys. Suppl. Ser.\/} {\bf #1}, #2}
\def \AJ #1 #2 {{\em Astron. J.\/} {\bf #1}, #2}
\def \ANNREV #1 #2 {{\em Ann. Rev. Astron. Astrophys.\/} {\bf #1}, #2}
\def \APJ #1 #2 {{\em Astrophys. J.\/} {\bf #1}, #2}
\def \APJL #1 #2 {{\em Astrophys. J. Lett.\/} {\bf #1}, L#2}
\def \APJS #1 #2 {{\em Astrophys. J. Suppl.\/} {\bf #1}, #2}
\def \APSS #1 #2 {{\em Astrophys. Space Sci.\/} {\bf #1}, #2}
\def \ASR #1 #2 {{\em Adv. Space Res.\/} {\bf #1}, #2}
\def \BAIC #1 #2 {{\em Bull. Astron. Inst. Czechosl.\/} {\bf #1}, #2}
\def \JSQRT #1 #2 {{\em J. Quant. Spectrosc. Radiat. Transfer\/} {\bf #1}, #2}
\def \MN #1 #2 {{\em Mon. Not. R. Astr. Soc.\/} {\bf #1}, #2}
\def \MEM #1 #2 {{\em Mem. R. Astr. Soc.\/} {\bf #1}, #2}
\def \PLR #1 #2 {{\em Phys. Lett. Rev.\/} {\bf #1}, #2}
\def \PASJ #1 #2 {{\em Publ. Astron. Soc. Japan\/} {\bf #1}, #2}
\def \PASP #1 #2 {{\em Publ. Astr. Soc. Pacific\/} {\bf #1}, #2}
\def \NAT #1 #2 {{\em Nature\/} {\bf #1}, #2}
\def \SAIT #1 #2 {{\em Mem.\ Soc.\ Astron.\ It.\/} {\bf #1}, #2}
\def \MESS #1 #2 {{\em The Messenger\/} {\bf #1}, #2}
\def \ASTRNACH #1 #2 {{\em Astron. Nach.\/} {\bf #1}, #2}
\def \AGPSR #1 #2 {{\em ASI Special Publication\/} {\bf #1}, #2}
\def\b12{{B_{12}}}

\def\i45{{I_{45}}}
\def\be{\begin{eqnarray}}
\def\ee{\end{eqnarray}}

\begin{opening}

\title{Gamma-Ray Pulsars: Modeling and Searches}
\author{M. A. McLaughlin$^{1}$ \& J. M. Cordes$^{2}$}
\institute{$^1$Jodrell Bank Observatory, University of Manchester, Macclesfield, Cheshire, SK11 9DL, UK\\ $^2$Astronomy Department and NAIC, Cornell University, Ithaca, NY 14853, USA}
\date{} 
\end{opening}

\begin{document}

\oddpagefooter{}{}{} 
\evenpagefooter{}{}{} 
\medskip  

\begin{abstract} 
Using a likelihood analysis 
and EGRET detections, upper limits and diffuse background measurements,
we find a best-fit luminosity law $L \propto P^{-1.7}B^{1.2}$ for the gamma-ray pulsar population.
We find that roughly 30 of the 170 
unidentified EGRET sources are likely to be pulsars. This is roughly twice the number of
known radio pulsars which are plausibly associated with
unidentified EGRET sources. We predict that AGILE will  
detect roughly 70 pulsars as point sources, including 12 which will be able
to be detected in blind periodicity searches.
GLAST should detect roughly 1200 pulsars (including only 200 currently known radio pulsars),
210 of which will be able
to be detected in blind searches. 
 We discuss methods of searching for pulsars in gamma-ray data
and present results from our searches for gamma-ray periodicities from new radio pulsars
associated with unidentified EGRET sources.
\end{abstract}

\medskip

\section{Introduction}

Because pulsars emit the majority of their
spin-down energy in gamma-rays, understanding their emission at these energies
is crucial for
forming a complete picture of pulsar energetics. However, while the radio pulsar population now numbers almost 1500, pulsed gamma rays have been
detected from less than 10 sources. With these sparse statistics, addressing important issues such
as how gamma-ray luminosity depends on spin-down parameters, the gamma-ray pulsar emission mechanism, the relationship between radio and gamma-ray beams,
the pulsar contribution to the unidentified EGRET source population, and the pulsar science prospects
of future gamma-ray  missions is difficult.
For this reason, we developed a likelihood analysis which uses the EGRET pulsar
detections, upper limits, and diffuse background measurements to characterize some properties of
the gamma-ray pulsar population. In this paper, we outline our updates to the analysis of
McLaughlin \& Cordes 2000 (hereafter MC00),
 present the new results, and discuss the prospects of AGILE and
GLAST for pulsar science.
We also discuss issues involved in searching for gamma-ray pulsars
and present the methodology and results
from searches for gamma-rays from new radio pulsars.
Such searches are difficult due to the sparseness of gamma-ray photons and the lack of
contemporaneous radio ephemerides.

\section{Model}

Our likelihood function is a product of the individual likelihoods
for pulsar detections, upper limits and diffuse background measurements. We
model a pulsar's gamma-ray luminosity $L$ as a power-law in period $P$ (in seconds) and
magnetic field $\b12$ (in units of $10^{12}$ G) such that $L = \gamma P^{-\alpha}\b12^{\beta}$. 
We assume a spin-down
law with braking index $n$ to calculate a population-averaged luminosity. 
To calculate all likelihoods we assume a broad
beaming solid angle of $2\pi$.
To calculate the diffuse background likelihood, we assume
a constant pulsar birthrate of 1/100 yr, a Galactic age of
$10^{10}$ years, and a maximum gamma-ray efficiency of 1/2. We allow pulsars to
contribute up to 10\% of the total diffuse flux and assume that they are distributed in a
Gaussian disk of scale 6~kpc with exponential halo of scale 0.5~kpc and molecular ring at
4~kpc with width 1.5~kpc.
Please see MC00 for detailed descriptions of the likelihood calculations.

In MC00, we used the Taylor \& Cordes (1993) electron density model to calculate pulsar distances.
In this analysis, we use Cordes \& Lazio (2002). This has not changed 
the results dramatically, but has been important for some individual objects. For example,
the implied gamma-ray efficiency for B1055$-$52 has decreased from 30\% to 5\%. In MC00, we
assumed a single value for the population's magnetic field $B$ and initial spin period $P_0$.
In this analysis, we
fit for a lognormal distribution for $B$ characterized by mean $<B>$ and rms $\sigma_B$
and a flat distribution for $P_0$ bounded by $P_{0,min}$ and $P_{0,max}$.

\section{Results}

Our EGRET data (described in MC00) consist of 7 pulsar detections, 353 pulsed flux upper limits
and 3 diffuse background measurements. 
We calculate the likelihood for a range of values for 
$\alpha$, $\beta$, $\gamma$, $n$,  $<B>$,  $\sigma_B$, $P_{0,min}$, and $P_{0,max}$ and find
well-defined maxima in $\alpha$, $\beta$, and $\gamma$ of 1.7, 1.2, and 32.4. We cannot constrain
$n$, $<B>$,  $\sigma_B$, $P_{0,min}$, or $P_{0,max}$ as for
any reasonable values the pulsar contribution to the diffuse flux
is $\ll$~10\%. 
As the best-fit luminosity law therefore only depends
on detections and upper limits,
it is independent of all assumptions aside from beaming solid angle and distance.
Table I shows that our best-fit law
is most similar to ones in which the luminosity is
proportional to the voltage drop across the polar cap. It is not consistent with 
luminosity proportional to spin-down energy and is also quite different from the
best-fit OSSE (McLaughlin et al. 2000) and radio (Arzoumanian et al. 2002) laws.

\begin{table}[h]
\caption{Comparison with Luminosity Laws}
\begin{center}
\begin{tabular}{lll}
\hline
\multicolumn{1}{c}{Law} & \multicolumn{1}{c}{$P$, $B$ Dependence} \\
\hline
EGRET best fit		&	$P^{-1.7}B^{1.2}$	\\
$L \propto \Delta V$    &       $P^{-2}B$       \\
$L \propto \dot{E}$	&	$P^{-4}B^{2}$	\\
OSSE best fit	&	$P^{-8.3}B^{7.6}$	\\
Radio best fit	&	$P^{-1.3}B^{0.4}$	\\
\hline
\hline
\end{tabular}
\end{center}
\end{table}

We can use the best-fit law and the assumptions of Section 2 to calculate the gamma-ray flux
distribution for a model population of pulsars (see Figure~1).
To do this, we must
assume values for $n$ and for the $B$ and $P_0$ distributions.  
For only five pulsars has  $n$ been measured, 
with values ranging from 1.5 to 3 (see Zhang et al. 2001 and references therein). 
We adopt $n = 2.5$, as measured
reliably for the Crab (Lyne et al. 1993).
Fitting a lognormal
distribution to the magnetic fields (i.e. $\b12^{2} = 10^{12} P \dot{P}$) of 1244 non-recycled 
pulsars, we find a mean of 12.1~G and rms of 0.65~G.
$P_0$ has
only be estimated for eight pulsars associated with supernova remnants with independent age
estimates and/or measurable pulsar proper motions. With estimates ranging from $< 14$~ms for
J0537$-$6910 to 139~ms for J0538+2817 (see Migliazzo et al. 2002, Kramer et al. 2003, and references therein),
we adopt a flat distribution from 10 to 150~ms.

\begin{figure}
\centerline{\psfig{figure=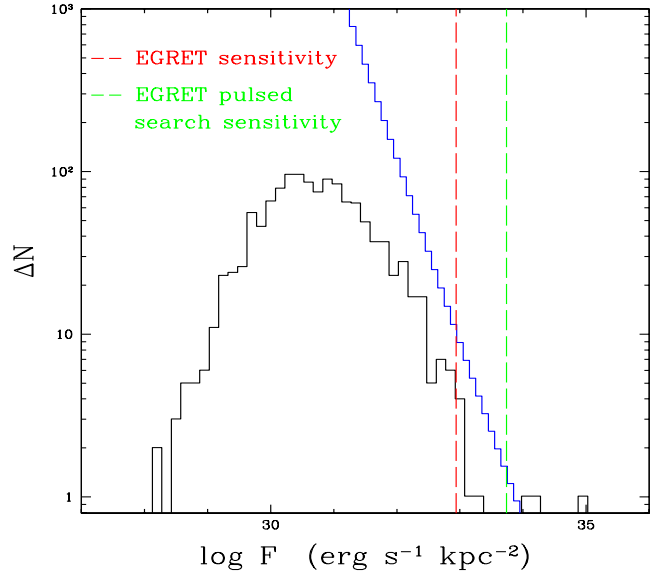,height=8cm,width=9.cm,angle=0}}
\caption[h]{The black line shows predicted fluxes for all 1244 non-recycled pulsars with measured $P$ 
and $\dot{P}$. The blue line
shows the predicted flux histogram for a model population of pulsars. Red and green lines show EGRET's
sensitivity to
point and pulsed sources.}
\end{figure}

This analysis shows that, given EGRET's point source sensitivity of
$10^{-7}$~ph~cm$^{-2}~$s$^{-1}$ $\sim 10^{32.9}$~ergs s$^{-1}$ kpc$^{-2}$ (for a 
spectral index of $-2$), EGRET should
have detected roughly 40 pulsars as point sources. For a pulsed-search sensitivity of
$6\times10^{-7}$~ph~cm$^{-2}$s$^{-1}$ $\sim 10^{33.7}$~ergs s$^{-1}$ kpc$^{-2}$ (for a
2-week integration, a duty cycle of 1/2, and a minimum $N_{s}^{2}/N_{t}$ of 50, where
$N_s$ is the number of source counts and $N_t$ is the number of total counts), 
EGRET could have detected six pulsars in blind searches. 
Of the 170 unidentified EGRET sources,
73 are within 10~degrees of the Galactic plane and 47 are non-variable (McLaughlin 2001),
suggestive of a pulsar
origin and consistent with our estimate of 40 pulsars detected as point sources.
We can 
also compare with the analysis of Kramer et al. (2003) which found that,
of the 48 positional coincidences between radio pulsars
and unidentified EGRET sources, 
$19 \pm 6$ are likely to be real.
The estimate of 40 EGRET pulsars is dependent
upon a beaming solid angle of $2\pi$. For a beaming solid angle of $\pi$, EGRET should have 
detected 20 pulsars as point sources. However, given the analysis of Kramer et al., this would
imply that  all pulsars detected by EGRET are currently known radio pulsars,
a hypothesis which seems unlikely given that
we know of at least one gamma-ray pulsar (Geminga) which is  not detected at radio wavelengths.

In Table II, we list all radio pulsars with high model-predicted
fluxes that are associated with unidentified EGRET sources\footnote{J1747--2958 actually lies
outside of the 95\% position confidence contour of 3EG~J1746--2851. However, because of diffuse model and
positional uncertainties, especially towards the Galactic center, we believe this is a possible association.}
. Six of these 12 pulsars have
been detected in the Parkes Multibeam Pulsar Survey, which has discovered half of the 80
radio pulsars with ages $<$ 100~kyr. We also list the EGRET fluxes (in
units of $10^{33}$ ergs~s$^{-1}$~kpc$^{-2}$) and
variability indices  (McLaughlin 2001)
of these sources. While all of the known EGRET pulsars have
non-variable fluxes, some of these unidentified EGRET sources appear to be highly
variable. It is difficult to determine whether these are spurious coincidences or if there is
indeed some method through which pulsars may produce
variable gamma-ray emission. One method is through searching for gamma-ray pulsations (see Section~4).

\begin{table}[h]
\caption{Plausible Associations Between Pulsars and Unidentified EGRET Sources}
\begin{center}
\begin{tabular}{lllll}
\hline
\multicolumn{1}{c}{Pulsar} & \multicolumn{1}{c}{3EG Source} & \multicolumn{1}{c}{$F_{\gamma}$} & \multicolumn{1}{c}{$V$} & $N_{vp}$ \\
\hline
J1747$-$2958	&	3EG J1746$-$2851	&	10.7& 2.80 & 12	\\
J1637$-$4642    &       3EG J1639$-$4702        &       8.8     &       1.88 & 5    \\
J1413$-$6141    &       3EG J1410$-$6147        &       7.5     &       0.72   & 4 \\
B1823$-$13      &       3EG J1826$-$1302        &       6.6     &       5.89   & 4 \\
J1420$-$6048    &       3EG J1420$-$6038        &       6.2     &       1.14   & 5 \\
B1853+01	&	3EG J1856+0114	&	6.2	&	1.32	& 2 \\
J1837$-$0604    &       3EG J1837$-$0606        &       6.0     &       2.81 &4    \\
J1015$-$5719    &       3EG J1014$-$5705        &       5.5     &       0.55   & 7 \\
J2021+3651	&	3EG J2021+3716	&	5.4	&	2.5	& 8	\\
J1105$-$6107    &       3EG J1102$-$6103        &       5.2     &       2.38	&7    \\
J1016$-$5857    &       3EG J1013$-$5915        &       5.0     &       0.15   &7 \\
J2229+6114	&	3EG J2227+6122	&	4.4	&	0.19	& 2\\
\hline
\end{tabular}
\end{center}
\end{table}

Given the expected sensitivities of AGILE and GLAST, we can 
predict the pulsar populations that each instrument will see. In Table III, we list 
the sensitivities (in units of $10^{31}$ ergs~s$^{-1}$~kpc$^{-2}$), and the numbers of
expected detections of pulsars as point sources and in blind periodicity searches.
In column two, the number of known radio pulsars that are expected to be detected is
listed in parentheses (i.e. for EGRET, our model predicts that 17 out of the 40 total
pulsar detections are
known radio pulsars). In column four, the number in parentheses lists how
many EGRET unidentified sources could be identified as pulsars in a blind periodicity search.
These are optimistic estimates as they assume a 2-week, on-axis exposure with fresh gas.  
Note that GLAST will be able to detect pulsations from ALL of the
unidentified sources in blind periodicity searches, unambiguously determining their nature.
All of the estimates in Table III depend on a beaming solid angle
of $2\pi$ and will decrease linearly with decreasing solid angle.

\begin{table}[h]
\caption{Number of Expected Detections for Various Instruments}
\begin{center}
\begin{tabular}{lllll}
\hline
\multicolumn{1}{c}{Instrument} & \multicolumn{1}{c}{Point $S_{min}$} & \multicolumn{1}{c}{Point Detections} & \multicolumn{1}{c}{Pulsed  $S_{min}$} & \multicolumn{1}{c}{Pulsed Detections} \\
\hline
EGRET	& 	90	&		40(17)	&	550	& 6(10)		\\
AGILE	&	45	&	70(25)	&	275	& 12(52)		\\
GLAST	&	3	&	1200(200)	& 25	&	210(170) \\
\hline
\end{tabular}
\end{center}
\end{table}

\section{Gamma-Ray Pulsar Searching}

With many pulsars expected to be detected at gamma-ray energies but not at radio wavelengths,
developing methods to search for gamma-ray periodicities is crucial. We are applying such
methods to search EGRET data for gamma-rays from the new radio
pulsars listed in Table~II.
While we have current radio ephemerides for these objects, they are too imprecise 
to allow  extrapolation back to the time the data were taken. Therefore,
searching over a wide range of frequency $f$ and frequency derivative $\dot{f}$,
centered around those predicted by the current ephemeris, is necessary. To calculate
the range to search, we assume errors in $f$ and $\dot{f}$ of roughly
$10^{-8}$~s$^{-1}$ and $10^{-12}$~s$^{-2}$, typically many  times greater than the quoted timing errors,
to account for glitches and timing noise.
We space trial values of $f$ and $\dot{f}$ so
that the smearing in the final folded profile is $< 10$\% of the
period. For each trial, we barycenter all photons to the pulsar's position and fold all those
with energies $E_\gamma \ge 100$~MeV within
an energy-dependent acceptance cone of radius $\theta \le 5^{o}.85(E_{\gamma}/100 {\rm MeV})^{-0.534}$,
weighting photons by the PSF (e.g. Ramanamurthy et al. 1996).
We compute the significance of each profile
 using $\chi^{2}$,  $H$ (e.g. de Jager 1994), and $Z^{2}$ (e.g. Buccheri et al. 1983),
normalizing by the number of trials,
and inspect the most significant signals 
as determined by all of those three methods.

We have applied this method to all of the pulsars listed in Table~II 
and to the known EGRET pulsars B0531+21
and B1706--44.
 For these known pulsars, we use ephemerides based on only 2~yrs of
radio data for comparison with the timing solutions of similar timescale for the new pulsars.
Examining the known pulsar results show that our inclusion of
all photons above 100~MeV is optimal for detection; a lower cutoff results in less significant
detections due to increased background photons but higher cutoffs result in 
too few pulsar photons. We find that for both known pulsars and in all viewing periods (VPs),
weighting by the PSF increases the significance of detections. 
We find that $\chi^{2}$, $H$, and $Z^{2}$ generally return the same values for best $f$ and $\dot{f}$ 
but that both $H$ and $Z^{2}$ return incorrect values for B0531+21 in some VPs where the
pulsar is weak.

For each pulsar in Table~II, we have searched all VPs for which the pulsar
is less than 10$^{o}$ off-axis, with the number searched given in Table~II.
Using $f$ and $\dot{f}$ errors as quoted above, we only detect B0531+21
in the last seven of 14 VPs. Widening the range of $f$ and $\dot{f}$ searched, we detect
the pulsar in all VPs. While B0531+21 is strong enough that it is still the most significant signal
over the wider range, this would not be the case for the new pulsars.
We detect B1706--44 in two of 
four VPs. If we do not use PSF weighting, only one of these detections is significant. 
Generally, detection significance for the B0531+21 and B1706--44 trials varies
as expected with exposure time and off-axis angle, but we do not detect B1706--44 in one viewing period
which has similar exposure and off-axis angle to another in which it is detected with high significance.
 For comparison, J1706--44 has a 
a flux just less than that of 3EG~J1746--2851 but higher than those of the other unidentified sources
listed in Table~II.

From two of the Table~II pulsars, 
 we detect tantalizing signals.
In three 
of the 14 VPs searched for J1747$-$2958, the highest significance
periodicities have similar pulse profile shapes (see Figure~2). This unique shape is not
seen in any observations of any other pulsars. While the component
amplitude ratio and widths vary for the different VPs, we see similar variations 
 for the B0531+21
detections. It is unclear why this periodicity
is not seen in any of the other VPs.
The timing model may not be accurate enough to extrapolate to the  earliest VPs, but there
are some VPs between the first and second detections in which we would expect to
see the signal but do not.
This may be due to variations in the flux of the source.
Similarly, in two of the eight VPs analyzed for J2021+3651,
we detect high significance periodicities
with similar profile shapes (see Figure~2). Again, why we do not detect the pulsar in other
VPs is unclear. While we are working on more thorough analyses of these pulsars, determining
whether they are in fact gamma-ray pulsars may have to wait for future observations with a current
radio ephemeris. From several other pulsars in Table~II we detect marginally significant
periodicities. However, none of the profile shapes are repeatable across VPs. 

\begin{figure}
\centerline{\psfig{figure=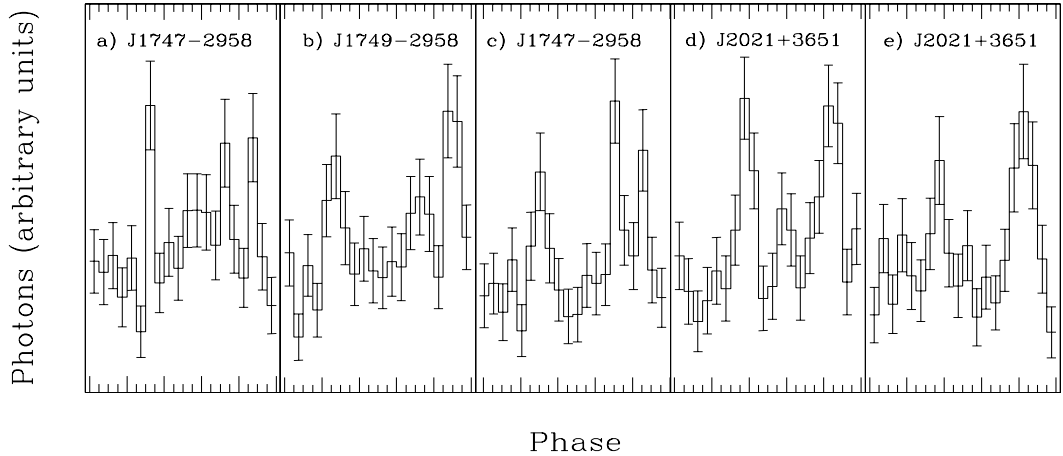,height=6cm,width=14.cm,angle=0}}
\caption[h]{Folded profiles, aligned to show shape similarities.
 Table~IV lists the VPs,
dates, observation lengths (in days), number of photons, and off-axis angle (in degrees) for each detection.}
\end{figure}

\begin{table}[h]
\caption{Possible Pulsar Detections}
\begin{center}
\begin{tabular}{ccccccc}
\hline
\multicolumn{1}{c}{} & \multicolumn{1}{c}{Pulsar} & \multicolumn{1}{c}{Viewing Period} & \multicolumn{1}{c}{Date} & Length & $N_p$ & Off-Axis Angle  \\
\hline
a)	& J1747$-$2958	&	4210	& Jun 1995 & 6.8 &  2395 & 4.0\\
b)	& J1747$-$2958   & 5080 & Dec 1995 &  7.2 & 2194 & 7.2 \\
c) 	& J1747$-$2958   & 6250 & Aug 1997 &  14.0 &  3597 & 2.3 \\
d)	& J2021+3651	& 0020 & Jun 1991 & 8.2 & 8902 & 3.1	\\
e)	& J2021+3651    & 3181 & Feb 1994 & 7.0 & 1934 &6.8	\\
\hline
\end{tabular}
\end{center}
\end{table}

\section{Future}

There are several improvements that we would like to make to our current population model. 
Obviously, all pulsars will not beam towards us with the same solid angle.
While we believe that this assumption is sufficient for making the large-scale predictions of
Table III, it does not allow us to explain the detection or non-detection of individual
objects.
We would also like to incorporate a more realistic pulsar spatial distribution,
incorporating spiral arms. This will not make a large difference for the EGRET and AGILE
detection statistics but may be important for GLAST, which will probe the population of
weak, distant pulsars. Our model currently ignores gamma-ray emission from millisecond
pulsars. But with the Kuiper et al. (2000) announcement of a probable detection of pulsed gamma rays from
pulsar J0218+4232, 
this issue must
be understood, as it could be important for the detectable populations of
AGILE and GLAST. Finally, before the launch of GLAST and AGILE, we must further optimize
pulsar search techniques so that we determine the most efficient ways of performing blind
searches for gamma-ray pulsars on these data.

\acknowledgements
We thank Fernando Camilo and Mallory Roberts for providing up-to-date radio ephemerides.

\end{document}